\documentclass[prc,amsmath,twocolumn,showpacs,superscriptaddress]{revtex4}
\usepackage[dvips]{graphicx}

\begin{document}

\title{Extracting (n,$\gamma$) direct capture cross sections from Coulomb dissociation: \\
application to $^{14}$C(n,$\gamma$)$^{15}$C}

\author{N.~C.~Summers}
 \affiliation{LLNL, P.O.~Box 808, L-414, Livermore, California 94551}
 \affiliation{National Superconducting Cyclotron Laboratory, Michigan State University, East Lansing, Michigan 48824}
 \author{F.~M.~Nunes}
 \affiliation{National Superconducting Cyclotron Laboratory, Michigan State University, East Lansing, Michigan 48824}
 \affiliation{Department of Physics and Astronomy, Michigan State University, East Lansing, Michigan 48824}

\date{\today}

\begin{abstract}
A methodology for extracting neutron direct capture rates from Coulomb Dissociation data
is developed and applied to the Coulomb dissociation of $^{15}$C on $^{208}$Pb at $68$ MeV/nucleon.
Full Continuum Discretized Coupled Channel calculations are performed and an asymptotic normalization
coefficient is determined from a fit to the breakup data. Direct neutron capture calculations using
the extracted asymptotic normalization coefficient provide $(n,\gamma)$ cross sections
consistent with direct measurements. Our results show that the Coulomb Dissociation data
can be reliably used for extracting the cross section for $^{14}$C(n,$\gamma$)$^{15}$C if
the appropriate reaction theory is used. The resulting error bars are of comparable magnitude
to those from the direct measurement. This procedure can be used more generally 
to extract capture cross sections from breakup reactions whenever the 
desired capture process is fully peripheral.
\end{abstract}


\maketitle

\emph{Introduction}.
Neutron direct capture reactions can play an important role in a
variety of astrophysical sites, including big bang, stellar
environments, and supernovae. Often, at the relevant energies,
these capture rates are extremely small, making the direct
measurement very difficult. Presently, direct measurements of
neutron captures on short lived unstable nuclei are not possible.
The Coulomb dissociation method \cite{baur86} provides an
alternative option with different systematic errors and
theoretical challenges. The aim of this work is to provide a
methodical and reliable procedure to extract the desired
$(n,\gamma)$ cross section from intermediate energy Coulomb
Dissociation data.

Here we consider the particular example of $^{14}$C(n,$\gamma$)$^{15}$C. This reaction is of
interest to astrophysics for a variety of reasons: (i) it is the slowest
in the neutron induced CNO cycle that takes places in AGB stars \cite{cno}; (ii) it has impact
on the abundances of the heavy elements produced by non-homogeneous big bang models \cite{bigbang}; (iii) it modifies the abundances resulting from the r-process in massive Type II supernovae
\cite{rprocess}.
In addition, it is one of the few that has been repeatedly studied through
Coulomb Dissociation and for which direct measurements exist.

Shortly after its astrophysical relevance was identified, a first
measurement of $^{14}$C(n,$\gamma$)$^{15}$C was performed at
Karlsruhe \cite{beer92}. As the container surrounding the target
had been strongly activated by a previous experiment, the
measurement had to be repeated many years later \cite{reifarth05}
in order to obtain a reliable result. In the meantime, a number of
Coulomb dissociation experiments have been performed using a
$^{15}$C beam on  a $^{208}$Pb target, at several beam energies:
the lowest data is available at 35 MeV/nucleon \cite{horvath02},
an intermediate energy measurement was performed at 68 MeV/nucleon
\cite{nakamura03} and a high energy experiment at 600 MeV/nucleon
\cite{pramanik03}. Meanwhile, another indirect method to obtain
the (n,$\gamma$) rate based on mirror symmetry \cite{timofeyuk06}
showed discrepancies between all three methods: the direct method,
the indirect Coulomb dissociation method and the mirror symmetry
method. Then recently, the new direct measurements by Reifarth
were revised and published in a conference proceeding
\cite{reifarth06}, which agreed with the Coulomb dissociation
measurements from RIKEN \cite{nakamura03} and the mirror symmetry
results \cite{timofeyuk06}. The discrepancy with the MSU data
\cite{horvath02} still remained.

Due to the weak binding of the $^{15}$C ground state, and because
there are no low lying resonances, the $(n,\gamma)$ cross section
is mainly determined through the E1 direct transition from an
initial p-wave scattering state to the ground state, with a small
branch to the only excited bound state \cite{timofeyuk06}. It is
well known that this dominant E1 transition is completely
peripheral and thus essentially fixed by the asymptotic
normalization coefficient (ANC) of $\langle ^{14}$C$|
^{15}$C$\rangle$ \cite{timofeyuk06}. The $1/2^+$ ground state is a
halo n$-^{14}$C s-wave, and the $5/2^+$ excited state is mostly
d-wave. In both cases the relevant physics can be described by a
single particle model. As the continuum is structureless below 1
MeV, the capture cross section in this low energy region is not
very dependent on the details of the scattering potential chosen
--- such as the radius and diffuseness --- but is strongly
dependent on the ANC of the bound state wavefunction.

While the calculation of the neutron capture rate is straightforward, the Coulomb
dissociation requires a reliable reaction model. All three Coulomb dissociation
measurements were analyzed with the virtual photon method \cite{bertulani}, which is based on
a first-order Coulomb-only semiclassical theory \cite{alder}. In \cite{horvath02}, the nuclear
component was determined from the scaling of the breakup cross section on
lighter targets, and subsequently subtracted from the total measured cross section.
In \cite{nakamura03}, the nuclear was excluded through the selection of impact parameters
larger than $b >30$ bm. Once a Coulomb only cross section is obtained, the virtual photon method
provides a simple proportionality relation to extract the desired $(n,\gamma)$ cross section.

The continuum discretized coupled channel method (CDCC) \cite{cdcc} provides a non-perturbative
framework in which to describe the breakup process, treating Coulomb and nuclear effects on
the same footing. Multipole excitations are fully taken into account as well as final state
interaction effects (e.g. \cite{nunes}).
Systematic studies of Coulomb dissociation for loosely bound systems
on a variety of targets, spanning a range of beam energies, have shown that nuclear
scaling is not always reliable and nuclear-Coulomb interference can be very
large \cite{hussein}. In that work it is suggested that, rather than {\it massaging} the data
to obtain a {\it Coulomb only} cross section, CDCC be used as a standard
tool to analyze the full measured dissociation data.
In this frame of mind, one would start with a single particle structure model for the
projectile $^{15}$C$=n+^{14}$C and would adjust the parameters to obtain a good description of the
breakup data. The reaction model then already takes into account nuclear and Coulomb effects in a coherent manner.
That same potential model for the projectile which fits the breakup data would then provide the corresponding neutron capture cross section.

A potential model for the projectile describing the breakup cross
sections is not unique. Indeed, as shown in \cite{capel}, for
loosely bound systems, even if the geometries differ
significantly, it is essentially the ANC that determines the
normalization of the breakup cross section. This is also true for
the neutron capture cross section whenever it is completely
peripheral, as in the case of $^{14}$C(n,$\gamma$)$^{15}$C. 
The use of ANCs in the analysis of peripheral reactions was
first introduced for transfer processes \cite{xu94}. In
this work we propose to use the dissociation data to extract the
ANC of $^{15}$C(g.s.) and then obtain from it the neutron capture
cross section. We concentrate on the breakup data at 68
MeV/nucleon first, to show the reliability of our proposed
procedure, and then discuss difficulties in the analysis of the
data at 35 MeV/nucleon. We do not include the $600$ MeV/nucleon
experiment in our studies since, at these energies, converged CDCC
calculations are very computational intensive and, given that
relativistic effects are too strong to be considered
approximately, results would not be reliable. A study of the
theoretical sources of uncertainty in the procedure will also be
presented.

\emph{Methodology}.
In this work, we propose a procedure for extracting direct neutron capture rates for loosely bound
systems from Coulomb dissociation, independent of the specifics for the experiments. It relies on
the fact that both, the breakup cross section and the capture cross section are proportional to the
square of the ANC. The method is applied to $^{14}$C(n,$\gamma$)$^{15}$C that has been measured
directly but also through Coulomb dissociation.
\begin{table}
\begin{tabular}{cccc|cc}
 $V_s$  & $V_p$  & $R_0$ & $a$   &  $C_{s1/2}$ & Ref.              \\ \hline
  65.19 &  68.52 & 1.1   & 0.5   &      1.254  &                   \\
  61.17 &  64.96 & 1.15  & 0.5   &      1.272  & \cite{pramanik03} \\
  55.36 &  60.43 & 1.223 & 0.5   &      1.298  & \cite{terry04}    \\
  54.91 &  61.13 & 1.22  & 0.53  &      1.319  &                   \\
  54.23 &  61.65 & 1.22  & 0.56  &      1.342  &                   \\
  52.79 &  61.85 & 1.228 & 0.6   &      1.376  & \cite{capel03}    \\
\end{tabular}
\caption{\label{TABLE:pot}
$^{14}$C+n potential parameters and calculated ANCs.
The spin orbit depth was fixed at 5 MeV with the same radius and diffuseness as the central part.
}
\end{table}

Our starting point is a set of  $^{14}$C+n potentials which span a
range of ANCs. In Table~\ref{TABLE:pot} we compile some $^{14}$C+n
potentials from the literature and add a few of our own, all with
changing geometries. The $V_s$ potential strength is obtained from
fitting the binding energy of the s-wave ground state of $^{15}$C.
The same $V_s$ reproduces well the $^{15}$C d-wave resonance. The
$V_p$ potential strength is obtained from the binding energy of
the p$_{3/2}$ neutrons in $^{14}$C. $V_s$ ($V_p$) is used to
calculate the scattering states with even (odd) parity. For each
of these single particle potentials, CDCC calculations for the
breakup of $^{15}$C on $^{208}$Pb are performed. Both bound states
are included and fully coupled in the calculation. The CDCC model
space needed for convergence includes partial waves for the
internal motion of the projectile up to $l \le 3$ and maximum
$^{14}$C+n relative energy $E_{\mathrm{rel}}\le 8$ MeV, with a
fine discretization from $0-2$ MeV to obtain higher accuracy in
this region (the number of bins below 2 MeV in each partial wave
is $N_{s_{1/2}}=10, N_{p_{1/2}}=10, N_{p_{3/2}}=20, N_{d}=8,
N_{f}=5$). For the projectile-target relative motion, the maximum
partial wave included is $L_{\mathrm{max}}=9000$. The coupling
potentials are expanded in multipoles and we include all
multipolarities up to $Q=3$. As to the radial truncations, energy
bins are integrated out to $50$ fm and the CDCC equations are
solved up to $R_{\mathrm{max}}=1000$ fm.

For each of the breakup energy distribution theoretical curves,
the $\chi^2$ for the data of \cite{nakamura03} was calculated. The
function $\chi^2(C_{s_{1/2}})$ is quadratic and from its minimum
we determine the ANC $C_0$, with an error from
$\chi_{\mathrm{min}}^2+1$ \cite{pdg}. Based on this range of ANCs,
neutron capture cross sections are calculated with an error bar
originating from the fit to the breakup data. The code
\textsc{fresco} \cite{fresco} is used for both the breakup and
capture calculations.

In summary, the proposed procedure involves a few steps: (i)
define a set of neutron single particle potentials that generate a
range of ANCs; (ii) calculate the corresponding breakup energy
distributions in CDCC; (iii) for each single particle potential,
calculate the $\chi^2$ to the breakup data and fix the ANC $C_0$
by minimizing $\chi^2$ relative to the ANC; (iv) determine the
error $\epsilon_{C_0}$ in $C_0$ through $\chi_{\mathrm{min}}^2+1$;
(v) calculate the neutron capture cross section corresponding to
$C_0 \pm \epsilon_{C_0}$.

There are a few details that can be a source of uncertainty to
this procedure. As mentioned in the introduction, there is a
nuclear contribution to the breakup reaction which is included in
our CDCC calculations. Fragment-target optical potentials are
usually fitted to elastic scattering to reduce the ambiguity. In
this particular example we use the neutron optical potential from
Perey and Perey \cite{perey} and the $^{14}$C optical potential
was taken from $^{10}$Be+$^{208}$Pb scattering, as in
\cite{fukuda04}, with a mass scaled radius. It has been shown
\cite{capel03} that for the breakup of $^{15}$C on $^{208}$Pb at
68 MeV/nucleon, the sensitivity to details of these potentials is
small.

Finally there is the issue of the single particle structure of the
projectile. For other cases, the single particle approximation has
been shown to carry a significant error \cite{xcdcc}. As discussed
in previous works (see for example \cite{timofeyuk06} and Ref.\
within), the ground state of $^{15}$C is considered to be a very
good single particle case with spectroscopic factor very close to
$S_{s_{1/2}}=1$. Even if this were not the case, when both the low
energy capture and breakup are completely peripheral, the s-wave
component dominates the cross section. This part is directly
proportional to the bound state ANC, in such a way that the
dependence on the single particle parameters and the spectroscopic
factor is negligible once the ANC is fixed $C^2_{g.s.}=S_{s_{1/2}}
b^2_{2s_{1/2}}$. As mentioned above, in $^{15}$C there is also a
d-wave excited state. The contribution of this state to the
neutron capture is small and uncertainty in the structure of this
state does not affect the errors bars.

\begin{figure}
\includegraphics[width=7.5cm]{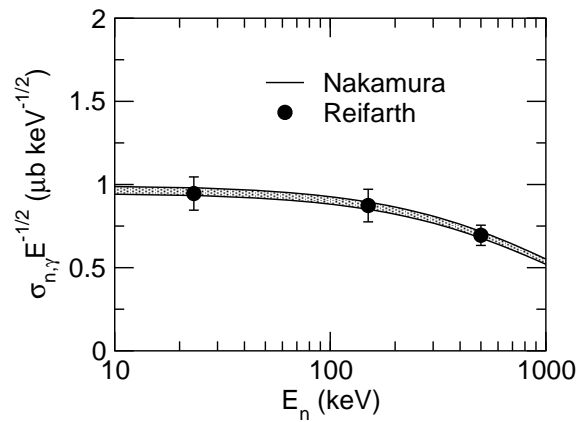}
\caption{\label{FIG:ng-riken} Capture cross sections, multiplied
by the energy factor $E^{-1/2}$, versus neutron energy. The shaded
area are cross sections obtained from the RIKEN data
\cite{nakamura03} and the black circles are the latest direct
measurements \cite{reifarth06}. }
\end{figure}

\emph{Results}.
\begin{figure}
\includegraphics[width=7.5cm]{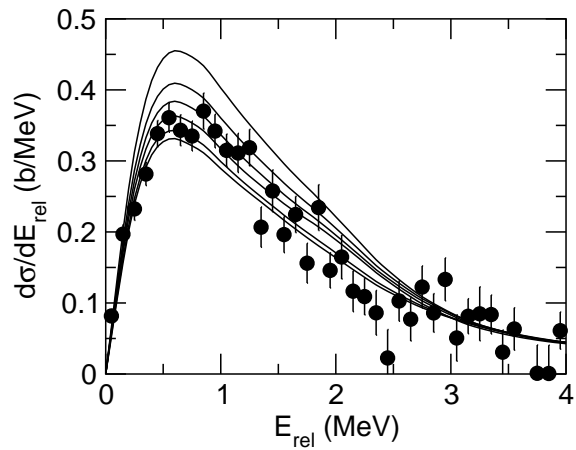}
\caption{\label{FIG:dsde-riken} Differential cross section with
respect to energy for $^{15}$C $\to$ $^{14}$C+n breakup on
$^{208}$Pb. The data are from Ref.~\cite{nakamura03} and the lines
represent the cross sections obtained from each of the potential
sets in Table~\ref{TABLE:pot}. }
\end{figure}
We present in Fig.~\ref{FIG:dsde-riken} the results for the
breakup cross sections of $^{15}$C on $^{208}$Pb at 68
MeV/nucleon, calculated within CDCC using the model space
described in the previous section. The shape of the distribution compares
well with the data. Most importantly, the peak of the cross
section scales linearly with the the ANC-squared. For each of the
theoretical curves, $\chi^2$ was calculated and a quadratic
relation with the ANC was determined. By minimizing the $\chi^2$,
an ANC was fixed $C_0=1.28\pm0.01$ fm$^{-1/2}$, the error bar
corresponding to $\chi_{\mathrm{min}}^2+1$. These allowed ANC
values produce a range of possible neutron capture cross sections,
shown by the shaded area in Fig.~\ref{FIG:ng-riken}.

Plotted in Fig.~\ref{FIG:ng-riken} is the range for
$\sigma_{n,\gamma} E^{-1/2}$ based on the RIKEN Coulomb
dissociation data, compared with the data from the latest direct
measurements \cite{reifarth06}. The agreement is very good. Note
that the lowest energy point in Fig.~\ref{FIG:ng-riken} at 23 keV
does not correspond to a mono-energetic neutron measurement. The
neutrons at this energy have a Maxwellian distribution, so an
averaged cross section is obtained. For 23 keV too, the prediction
based on the RIKEN Coulomb dissociation data ($7.0\pm0.2$ $\mu$b)
compares well with the direct measurement ($7.1\pm0.5$ $\mu$b).
For the purpose of the comparison in Fig.~\ref{FIG:ng-riken}, we
multiplied the 23 keV data point by 0.67, which is the factor one
obtains assuming a perfect E$^{-1/2}$ energy dependence in the
cross section (valid at this low energy).

A lower energy breakup measurement is also available
\cite{horvath02}. A direct comparison of theoretical cross
sections with the experimental data was not possible for this
experiment due to a non-linear energy response function of the
detectors. Therefore the theoretical cross sections had to be
folded with the detector efficiency in order to compare with the
data. The analysis in Ref.~\cite{horvath02} suggested an
$(n,\gamma)$ cross section approximately half that found in the
analysis on the RIKEN data presented in the previous section and
other direct and indirect measurements
\cite{reifarth05,pramanik03,timofeyuk06}. Here we present CDCC
calculations which test the assumptions that appeared in the
analysis of Ref.~\cite{horvath02}. For the purpose of this study
we use the single particle model of Ref.~\cite{terry04}, with
again the Perey-Perey \cite{perey} neutron-Pb potential and the
same optical potential for the core as in the previous section.

\begin{figure}
\includegraphics[width=7.5cm]{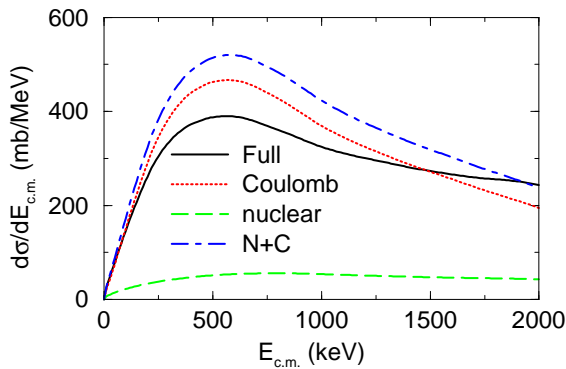}
\caption{\label{FIG:n+c} (Color online)
Nuclear and Coulomb interference in the Coulomb breakup of $^{15}$C at 35 MeV/nucleon.}
\end{figure}

The first important assumption in Ref.~\cite{horvath02} is that
the nuclear contribution can be subtracted from the data to leave
a \emph{Coulomb only} cross section. This was attempted by
measuring the breakup data on a range of targets from the heavy Pb
down to the light C target. Assumptions were made on how the
Coulomb and nuclear cross sections scale with target mass, and by
adding them incoherently, a least squares fit of the data was
performed to estimate the relative cross sections so that the
nuclear part could be subtracted.

In Fig.~\ref{FIG:n+c} we show the breakup energy distribution
for the full calculation (solid line), including both nuclear and Coulomb, Coulomb only
(dotted line), nuclear only (dashed line) and the incoherent sum of Coulomb and
nuclear (dot-dashed line). The nuclear contribution is not negligible and interference
effects are large, in agreement with the result of \cite{hussein}. Most importantly,
the shape of the distribution is changed when interference is taken into account.
At low energies the full calculation including nuclear and Coulomb coherently is actually less than the Coulomb only calculation.

\begin{figure}
\includegraphics[width=7.5cm]{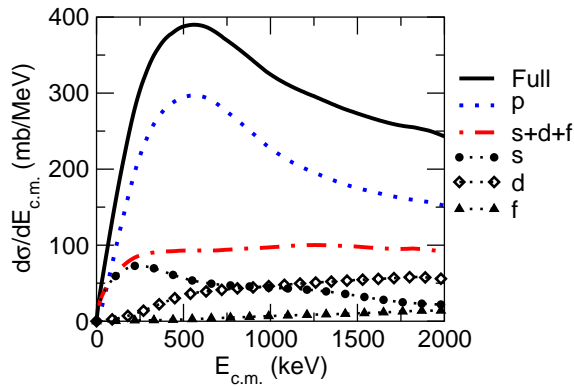}
\caption{\label{FIG:partial} (Color online)
Partial wave decomposition of the cross section in the Coulomb breakup of $^{15}$C at 35 MeV/nucleon.}
\end{figure}
The other main assumptions appear in the analysis of the detector efficiencies.
In order to calculate the efficiencies, a cross section must be entered into the Monte Carlo simulation of the detector response.
The cross section used in Ref.~\cite{horvath02} was obtained from first order perturbation theory.
At this lower energy, the main assumptions of this theory over simplify the reaction mechanism, namely that the breakup proceeds via a single step E1 transition.

In our CDCC calculations we have included all multipolarities up to $Q=3$ and $^{14}$C+n
partial waves up to $l \le 3$. In Fig.~\ref{FIG:partial} we show the contribution
of all the partial waves. Although the p-wave is dominant (and is mainly E1 but would contain some nuclear contributions too), one should not neglect the other partial waves as it affects the normalization and the shape of the distribution.

\begin{figure}
\includegraphics[width=7.5cm]{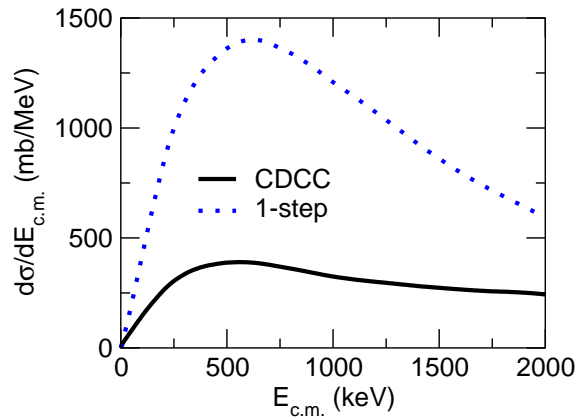}
\caption{\label{FIG:multistep} (Color online)
Higher-order effects in the Coulomb breakup of $^{15}$C at 35 MeV/nucleon.}
\end{figure}

In Fig.~\ref{FIG:multistep} we compare the full CDCC calculations with the corresponding one-step calculation. This corresponds to a DWBA calculation where
the optical potentials used are the CDCC coupling potentials as discussed
in \cite{nunes}. It had already been shown \cite{nunes} that multistep
effects can be very large at energies close to the Coulomb barrier.
Results presented in Fig.~\ref{FIG:multistep} show that at 35 MeV/nucleon multistep processes are also important and affect the normalization of the cross section.

These assumptions together overestimate the cross section used in
the detector simulations, which in turn underestimate the
efficiencies. Hence the theoretical cross sections which fit the
data, once folded with the detector response functions come out
much smaller than they should. This goes a long way to explaining
why the capture cross sections derived from this data are
approximately half that seen in other experiments. This outlines
the need for good theoretical-experimental communication beyond
the usual comparison at cross section level.

To summarize, in the analysis of \cite{horvath02} it was assumed that: (i) nuclear-Coulomb interference was insignificant; (ii) the outgoing neutrons were all in p-waves and (iii) the breakup process occurred in one-step. Given that these three assumptions are not correct, a re-analysis of this data would be highly desirable.

\emph{Conclusions}.
A procedure for determining the direct neutron capture reaction
from intermediate Coulomb dissociation data is presented. The
procedure is valid whenever the neutron capture is fully
peripheral. Full CDCC calculations for the breakup process are
compared to Coulomb dissociation data and a range of allowed
asymptotic normalization coefficients is extracted from $\chi^2$
minimization. Neutron capture cross sections consistent with this
range of ANCs are determined.

The method is applied to $^{14}$C(n,$\gamma$)$^{15}$C. The Coulomb
dissociation data of \cite{nakamura03} is analyzed with CDCC and
the ANC $C_0=1.28\pm0.01$ fm$^{-1/2}$ is obtained for the $^{15}$C
ground state. We show that the corresponding $(n,\gamma)$ cross
sections are consistent with direct measurements
\cite{reifarth06}, providing comparable accuracy, and consistent
with the results from mirror symmetry \cite{timofeyuk06}. Previous
discrepancies between all three methods where shown to be drastic
in Ref.~\cite{timofeyuk06}.
These discrepancies have been partially resolved by an improved
analysis of the direct measurement by Reifarth {\it et al.} \cite{reifarth06},
with the exception of \cite{horvath02}. Our calculations imply that
assumptions made in the analysis of Horv\'{a}th {\it et al.} \cite{horvath02}
are not valid. In order to include this measurement
in the extraction of the $^{14}$C(n,$\gamma$)$^{15}$C cross section,
a reanalysis of the data is needed.

\medskip

We acknowledge valuable discussions with M. Heil and R.Cyburt, and T.
Nakamura for providing the RIKEN Coulomb dissociation data.
This work was partially supported
by the Joint Institute for Nuclear Astrophysics at Michigan State
University (N.S.F. grant PHY0216783), the National Science Foundation
grant PHY-0555893, D.O.E grant DE-FG52-03NA00143
and in part performed under the auspices of the
U.S. Department of Energy by Lawrence Livermore National
Laboratory under Contract DE-AC52-07NA27344.


\end{document}